\documentclass[prb,twocolumn,showpacs,superscriptaddress]{revtex4}

\usepackage{graphicx}
\usepackage{dcolumn}

\usepackage{eucal}
\usepackage[dvips]{epsfig}
\usepackage{changebar}
\usepackage{amssymb}
\usepackage{amsfonts}

\usepackage{amsmath}

\newcommand{\BRA}[1]{\langle{#1}\vert}

\newcommand{\KET}[1]{\vert{#1}\rangle}

\newcommand{\EW}[1]{\left\langle{#1}\right\rangle}

\newcommand{\GREEN}[3]{\left\langle\hspace{-.4ex}\left\langle#1;
      #2\right\rangle\hspace{-.4ex}\right\rangle{#3}}

\begin{document}

\newcommand{\csp}{\;,\qquad\qquad} 
\title{Effects of dilution and disorder on magnetism in diluted spin systems}

\author{Guixin Tang}
 \email{tang@physik.hu-berlin.de}
\affiliation{Institut f\"ur Physik, Humboldt-Universit\"at zu Berlin,
 Newtonstra{\ss}e 15, D-12489 Berlin, Germany}
\affiliation{Department of Physics, Harbin Institute of
Technology, D-150001 Harbin, People's Republic of China}
\author{Wolfgang Nolting}
\affiliation{Institut f\"ur Physik, Humboldt-Universit\"at zu Berlin,
 Newtonstra{\ss}e 15, D-12489 Berlin, Germany}

\date{\today}

\begin{abstract}

The influence of configurational disorder on the magnetic properties of
diluted Heisenberg spin systems is studied with regard to the ferromagnetic
stability of diluted magnetic semiconductors.
The equation of motion of the magnon Green's function is decoupled by Tyablikov approximation.
With supercell approach, the concentrations of magnetic ions are determined by the size of the supercell in which
there is only one magnetic ion per supercell in our method.
In order to distinguish the influence of dilution and disorder, there are two kinds of supercells being used:
the \textit{diluted and ordered} case and the \textit{diluted and disordered} case.
The configurational averaging of magnon Green function due to disorder is treated in the augmented space formalism.
The random exchange integrals between two supercells are treated as a matrix.
The obtained magnon spectral densities are used to
calculate the temperature dependence of magnetization and Curie temperature.
The results are shown as following:
(i) dilution leads to increasing the averaged distance of two magnetic ions,
further decreases the effective exchange integrals and is main reason to reduce Curie temperature;
(ii) spatial position disorder of magnetic ions results in the dispersions of the exchange integrals between two supercells
and slightly changes ferromagnetic transition temperature;
(iii) the exponential damping of distance dependence obviously reduces Curie temperature
and should be set carefully in any phenomenological model.

\end{abstract}


\maketitle

\section{Introduction}
\label{sec:Introduction}

Following the discovery of a ferromagnetic transition at temperatures in excess of $100 K$, 
the diluted III-V magnetic semiconductors (DMS), which are realized by doping a semiconducting host 
material with low concentrations of magnetic impurities (typically manganese), 
have attracted a great deal of attention from both the experimental and theoretical point of view
due to their potential in spintronics applications
\cite{Matsukura98,Ohno98,Beschoten99,Konig00,Dietl01,Ohno01,Dietl02,DasSarma03,MacDonald05}. 

In DMS, low concentrations of magnetic impurities carrying localized magnetic moments (spins) form a diluted spin system. 
The random spatial distribution of the magnetic impurities breaks the translational symmetry of the crystal 
and thus greatly complicates the theoretical description of the material\cite{Priour04,DasSarma04,Hilbert0405,Bouzerar05}. 
Several different theoretical methods\cite{Bouzerar03,Chudnovskiy02,Berciu01,Berciu02}
have been performed to get the transition temperatures. 
In some stages of these calculations, disorder effects have been completely neglected or treated within 
the mean-field approximation (MFA). 
Monte Carlo (MC) simulations\cite{Mayr02,Kennett02,Calderon02,Zhou04,Bergqvist04,Bergqvist05,Sato04,Schliemann01} 
seem to provide a better way to include the positional disorder, but these theories usually assume classical spins. 
A proper treatment of the positional disorder of the localized moments and 
their quantum nature is necessary\cite{Hilbert0405,Schliemann03}. 
Recently, \textit{ab initio} calculations with supercell approaches 
\cite{JLXu05,Silva05,XYCui05,Raebiger05} 
are used to investigate the effects of disorder. 
Alternatively, the stochastic series expansion (SSE) quantum MC (QMC) method with $L\times L$ supercell 
is used to investigate the order-disorder transition in the diluted two-dimensional Heisenberg model 
with random site dilution\cite{Sandvik02}. 
But there are still doubts concerning the effect of disorder on magnetism. 

In fact, the effect of disorder on magnetism is an old and important problem in diluted spin systems\cite{Tahir-Kheli,Collection}, 
although only during the last few years have disorder effects in DMS been considered. 
In these early papers, a typical method is using coherent potential approximation (CPA) 
that is initially developed by Soven and Taylor\cite{Soven67,BEB} to treat the dynamics of a random Heisenberg-type Hamiltonian. 
However, most of these methods only include the short-range interaction and 
are hard to extend to the long-range exchange interaction. 
Up to now, lots of methods of investigating spin systems such as SSE-QMC\cite{Sandvik02} 
and density matrix renormalization group thoery\cite{Laflorencie06} only take into account short-range or nearest-neighbor interactions. 
But the experimental fact of the transition temperature with low concentrations implies that 
the exchange interaction between the localized moments is long-ranged in DMS\cite{Matsukura98,Beschoten99}. 
It means the long-range exchange interactions should be also included in the theoretical calculation 
of the magnetization and the transition temperature.   

Another method for calculating the properties of disordered systems is the augmented space formalism (ASF)\cite{Saha05,Alam04}, 
which is introduced by Mookerjee\cite{Mookerjee73} and centers around averaging functions of independent random variables. 
Rather than expanding the Green function in some manner and then averaging an appropriate set of terms, 
the random problem is transformed into an ordered one which is defined in a larger Hilbert space. 
This new Hilbert space is referred to as the augmented space, which may be described as the direct product of 
the Hilbert space spanned by the original Hamiltonian with a "disorder" space that describes the random variables. 
On transforming to this augmented space, a new nonrandom Hamiltonian can be defined such that 
configurational averages in real space for the random solid are equal to inner products 
in the augmented space\cite{Kaplan76,Kaplan80,Diehl76}. 

In this article, we present a new approach, which combines the supercell approach and the ASF, 
to study theoretically the influence of position disorder of magnetic ions on magnetization and 
the transition temperature in diluted spin systems on a disordered Heisenberg model. 
Firstly, the size of the supercell determines the concentration of magnetic ions in the host materials. 
There is one magnetic ion per supercell which, however, 
can only occupy the same site in a supercell (the \textit{diluted and ordered} case) or  
can occupy any site in a supercell (the \textit{diluted and disordered} case). 
In the \textit{diluted and disordered} case, 
the distance between two magnetic ions therefore becomes a random variable and then 
the effective Heisenberg exchange integrals, which are assumed to be a function of distance only, are random variables. 
In the framework of the ASF, the random exchange integrals are extended to matrices. 
Furthermore, the obtained spectral densities are used to calculate the temperature dependence of magnetization 
and Curie temperature of systems. 
Significantly, the spins are treated quantum mechanically in our approach 
although we use the supercell approach just like MC simulations. 
Moreover, the long-range exchange integrals are included. 
It should be mentioned here that the direct numerical diagonalization of the Green function\cite{Hilbert0405} 
is only applicable to the case of the finite size systems 
although it is a good method to include the long-range exchange integrals and to treat the spins quantum mechanically. 

The article is organized as follows. 
The theoretical methods are described in section \ref{sec:Model}. 
Section \ref{sec:Numerical_Studies} is concerned with the numerical studies and discussion. 
In section \ref{sec:Summary}, we conclude the article with a summary. 

\section{The model}
\label{sec:Model}

\begin{figure}[htb]
\centerline{\includegraphics[width=0.8\linewidth]{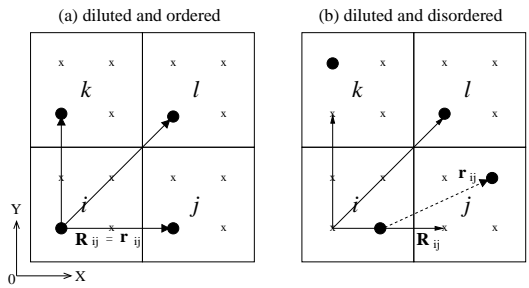}}
\caption{
\label{fig:supercell}
A two dimension view of the sc $\{2 \times 2 \times 1\}$ supercell ($x=25\%$) 
for (a) \textit{diluted and ordered} case, and (b) \textit{diluted and disordered} case. 
The small crosses represent the lattice sites and the black circles show the lattice sites occupied by magnetic ions. 
}
\end{figure}

For a diluted spin material $\mbox{A}_{1-x}\mbox{B}_{x}$ (A: non-magnetic; B: magnetic), 
the concentration $x$ ($x \in (0,1]$) of magnetic ion B is the ratio of the number of magnetic ions to all ions. 
If we consider a supercell that is built by non-magnetic ions with only \textbf{one} substituted by magnetic ions, 
the size of the supercell decides the concentration $x$. For example, for simple cubic (sc) systems, 
the concentration $x$ equals to $1 / (l \times m \times n)$ for the supercell with size $\{ l \times m \times n \}$, 
where $l$, $m$ and $n$ refer to the numbers of ions in $X$, $Y$ and $Z$ axis, respectively. 
In Fig. \ref{fig:supercell}, as an example, we give two kinds of possible realizations of 
a sc $\{2\times 2\times 1\}$ supercell that corresponds to the concentration $x=25\%$: 
(a) \textit{diluted and ordered} case, and (b) \textit{diluted and disordered} case. 
Each magnetic ion is located in the same lattice site of the supercell in (a), 
but in any possible site of the supercell in (b). 

To study the magnetic properties of diluted magnetic systems, 
we use the effective Heisenberg Hamiltonian
\begin{equation}
\label{eq:Heisenberg_Hamiltonian_1}
  H=-{\sum_{i,j=1}^{N}}J_{ij}\, \mathbf{S}_i\cdot \mathbf{S}_j
-\frac{1}{\hbar}g_J \mu_{\rm{B}} B {\sum_{i=1}^{N}} S_i^z\;.
\end{equation}
Here $i$ and $j$ label the sites occupied by the magnetic ions and
\mbox{$\mathbf{S}_i=\left(S_i^x,S_i^y,S_i^z\right)$} 
is the magnetic moment at lattice site $i$ with lattice vector $\mathbf{r}_i$. 
Moreover, in our supercell approximation, $i$ and $j$ can be also referred to the labels of the supercell 
because there is only one magnetic ion in each cell. 
The exchange parameter $J_{ij}$ is assumed to be a function of the distance 
$\vert \mathbf{r}_i - \mathbf{r}_j \vert$ only, independent of local environment. 
The Hamiltonian also contains a Zeeman coupling with external magnetic field \mbox{$\mathbf{B}=(0,0,B)$}.

Introducing the retarded magnon Green's function
\begin{equation}
\label{eq:Green_Function_Df}
  G_{ij}(E)=\GREEN{S_i^+}{S_j^-}{_E^{ret}}\;
\end{equation}
where $S_i^\pm=S_i^x\pm i S_i^y$, its equation of motion reads
\begin{multline}
\label{eq:Equation of Motion1}
  \left(E-g_J \mu_{\rm{B}} B\right)G_{ij}(E) =
  2\hbar^2\delta_{ij}\EW{S_i^z}
   - 2\hbar\sum_{m}J_{im} \\
\times \left(
  \GREEN{S_i^+ S_m^z}{S_j^-}{_E^{ret}}
  -\GREEN{S_m^+ S_i^z}{S_j^-}{_E^{ret}}\right)\;.
\end{multline}
To decouple the higher-order Green's function, we make the Tyablikov approximation 
on the rhs. of \eqref{eq:Equation of Motion1}. 
After rearranging, the equation of motion for $G_{ij}(E)$ is 
\begin{equation}
\label{eq:After_Tyablikov}
  \sum_{m} \left[\left( \omega - \sum_{n} J_{in}\right)\delta_{im}+ J_{im} \right]G_{mj}=\hbar\delta_{ij}\;,
\end{equation}
where $\omega=(E-g_J \mu_{\rm{B}} B)/(2\hbar\EW{S^z})$. 
Thus, the matrix form of Green's function can be expressed as:
\begin{equation}
\label{eq:GreenF_Matrix}
  \mathbf{G}(\omega)=\hbar \left(\omega\mathbf{I}-\mathbf{H}\right)^{-1} \;,
\end{equation}
where $\mathbf{I}$ is the identity matrix and 
the matrix $\mathbf{H}$ has the elements $H_{ij}=\delta_{ij}\sum_{n=1}^{N}J_{in}-J_{ij}$ 
which belong to the Hilbert space $\mathcal{H}$.

In the concentrated case, after Fourier transforming, 
one can evaluate the Green's function from its momentum representation: 
\begin{equation}
\label{eq:GreenF_Momentum}
  G(\mathbf{q},\omega)=\frac{\hbar} { \omega - \left[ J(\mathbf{0}) - J(\mathbf{q}) \right] } \;.
\end{equation}
Here $J(\mathbf{q}) = \sum_{I} J_{R_I} z_I \gamma_I(\mathbf{q})$ 
where $\sum_{I}$ corresponds to the summation over the $I$-th shell with a distance $R_{I}$ from a given site $0$, 
$z_I$ is the total number of sites in the $I$-th shell 
and $\gamma_I(\mathbf{q}) = \frac{1}{z_I} \sum_{\mathbf{r}^I_n} e^{-i \mathbf{r}^{I}_{n} \cdot \mathbf{q} }$ where 
the summation over $\mathbf{r}^{I}_{n}$ runs over each site located in the $I$-th shell. 

The equation (\ref{eq:GreenF_Momentum}) can also be applied to the calculation of the \textit{diluted and ordered} case if 
one considers all the distances $\mathbf{r}_{ij}$ between two magnetic ions as equal to 
the $\mathbf{R}_{ij}$ between the respective supercells. 
Due to the translational symmetry of the supercell, it is chosen as the Wigner-Seitz unit cell. 
The shell structure is determined by the spatial distribution of the supercells. 
For example, in Fig. \ref{fig:supercell}, 
if the supercell $i$ is set as the central supercell that refers to the $0$-th shell, 
the supercells $j$ and $k$ belong to the first shell and the supercell $l$ belongs to the second shell. 

But for the diluted spin system, equation (\ref{eq:GreenF_Momentum}) is not appropriate to be used directly 
because each lattice has different neighboring environment and the exchange integral between two supercells is different. 
The translational symmetry on lattice and supercell is absent. 
According to (b), the \textit{diluted and disordered} case of the Fig. \ref{fig:supercell}, 
the distance between two magnetic ions, which can occupy any site in their own supercell, 
is a random variable, e.g. $\mathbf{r}_{ij}$. 
Then all possible distances between two magnetic ions in different supercells are described as a set of independent random variables. 
For example, in the framework of the shell structure, the set of all possible distances between two magnetic ions, 
which belong to the $n$-th supercell of the $I$-th shell and the central supercell, 
is $\{r_1,r_2,...,r_k\}$ with the probability $\{c_{r_1},c_{r_2},...,c_{r_k}\}$ where $\sum_{i}{c_{r_i}}=1$, respectively. 
It leads to the exchange integrals $J_{R_{In}}$ represented as a random variables set 
$\{J_{r_1},J_{r_2},...,J_{r_k}\}$ with the probability density 
\begin{equation}
\label{eq:J_ij_dos}
p^k(J_{R_{In}}) = {\sum_{l=1}^{k}} c_{r_l} \delta(J_{R_{In}}-J_{r_l}) \;.
\end{equation}
It should be mentioned that, according to the shell structure, 
the set of exchange integrals $J_{R_{In}}$ and the probability density $p^k(J_{R_{In}})$ 
in the $I$-th shell are the same when $n$ runs over each supercell in the $I$-th shell. 
So, one simplifies them as $J_{R_I}$ and $p^k(J_{R_I})$. 

For each random variable $J_{R_I}$, according to the augmented space theorem, 
one can introduce a new Hilbert space $\phi^k$ such that $p^k(J_{R_I})$ corresponds to 
a suitably chosen operator $M_{I}^k$ on $\phi^k$ of rank $k$, spanned by $k$ possible values of $J_{R_I}$. 
If $\KET{f_{0}^{k}}$ (usually $\KET{1,0,\cdots,0}$) belongs to an orthonormal basis in $\phi^k$, then
\begin{equation}
p^k(J_{R_I})=-\frac{1}{\pi}\lim_{J \rightarrow J_{R_I}+i{0^+}} \mbox{Im}
  \langle f_{0}^{k} \vert (JI_k - M_{I}^k)^{-1} \vert f_{0}^{k} \rangle \;
\end{equation}
where $I_k$ is the identity operator on the space $\phi^k$. 
In other words, $\KET{f_{0}^{k}}$ and $M_{I}^k$ are chosen such that the spectral density of the operator $M_{I}^k$ 
with respect to $\KET{f_{0}^{k}}$ is the given probability distribution. 
With the suitable choice of the basis, the tridiagonal matrix representation of $M_{I}^k$ 
has $a_l$ ($l=1,\cdots,k$) down the diagonal and $b_m$ ($m=1,\cdots,k-1$) down the off-diagonal positions, 
which can be obtained by the continued-fraction expansion:
\begin{equation}
p^k(J_{R_I})=-\frac{1}{\pi}\lim_{J \rightarrow J_{R_I}+i{0^+}}\mbox{Im}\frac{1}{J-a_1-\frac{b_1^2}{J-a_2-\frac{b_2^2}{\cdots}}} \;.
\end{equation}
Thus the random exchange integral $J_{R_I}$ is transformed into the matrix representation $M_{I}^k$ in the Hilbert space $\phi^k$ 
where the randomness of the exchange parameters is completely included. 

Now, in the framework of the ASF, each supercell has the same spin term and 
the exchange integral from the central supercell to one special shell has the same matrix form. 
No supercell in the configurational distribution is distinguishable from any other. 
In other words, in the diluted spin system, the translational invariance that is destroyed on lattice is recovered on the supercell 
while the randomness is masked in the exchange integral matrix $M_{I}^k$. 
Note, the exchange integral matrix $M_{I}^k$ is the same within one and the same shell, but different among the different shells.
Here, we want to mention an interesting technique: the combinatorial method\cite{Florek02} 
that reduces initial eigenproblem to a less-dimensional one. 
In the case of finite systems, considering symmetry properties including methods of algebraic combinatorics, 
the Hamiltonian matrix is transformed to a block (quasidiagonal) form 
and the numerical solutions of eigenproblem are high precision and small resultant. 
In our approaches, due to recovering in the translational invariances, 
one can use the equation (\ref{eq:GreenF_Momentum}) to study an infinite system. 

In the ASF, the configurational averages in real space are replaced by 
inner products in the augmented space $\Sigma = \mathcal{H} \otimes \Phi$, where 
the "disorder" space $\Phi = \phi^1 \otimes \phi^2 \otimes \cdots \otimes \phi^I \otimes \cdots$ 
with $\phi^I \equiv \prod_1^{z_I} \otimes \phi^{k_I}$ being $z_I$ times the direct product of $\phi^{k_I}$ and 
$z_I$ is the total number of the supercells in the $I$-th shell. 
Moreover, $\KET{F_0} = \KET{F^1_{0}} \otimes \KET{F^2_{0}} \otimes \cdots \otimes \KET{F^I_{0}} \otimes \cdots$ 
is an orthonormal basis in the "disorder" space $\Phi$, 
where $\KET{F^I_{0}} \equiv \prod_1^{z_I} \otimes \KET{f_{0}^{k_I}}$. 
Note, in the expression of $\KET{f_{0}^{k_I}}$, we add the index $I$ to represent $\KET{f_{0}^{k}}$ in the $I$-shell. 
About the details of constructing the augmented space for a randomly disordered system with independent site-occupation variables, 
we refer the reader to a series of papers\cite{Mookerjee73,Kaplan76,Kaplan80,Diehl76}. 
The elements of Green function can be expressed as 
\begin{subequations}
\begin{eqnarray}
\label{eq:GreenF_ASF}
\mathbf{G}_{0n}(\omega) &=& \BRA{R_0 \otimes F_0} 
  \hbar \left( \omega\textbf{\~I}-\textbf{\~H} \right)^{-1} \KET{R_n \otimes F_0} \;, \\
\textbf{\~H} &=& \sum_{ij} \KET{R_i} \BRA{R_j} \otimes 
                 ( \delta_{ij} \sum_{m} \mathbf{J}_{im} - \mathbf{J}_{ij} ) \;.
\end{eqnarray}
\end{subequations}
Here, if $i$ refers to the central supercell and $j$ refers to the $n$-th supercell that is located in the $I$-th shell, then 
\begin{subequations}
\begin{eqnarray}
\mathbf{J}_{ij} &=& \mathbf{I}^1 \otimes \cdots \otimes \mathbf{I}^{I-1} \otimes \mathbf{M}^{I} \otimes \mathbf{I}^{I+1} \cdots \;, \\
\mathbf{M}^{I} &=& ( \prod_{1}^{n-1} \otimes \mathbf{I}_{k_I} )
                   \otimes M_{I}^{k_I} \otimes 
                   ( \prod_{n+1}^{z_I} \otimes \mathbf{I}_{k_I} ) \;, 
\end{eqnarray}
\end{subequations}
where $\prod_{m'}^{m} \otimes \mathbf{I}_{k_I}$ means $(m-m'+1)$ times the direct product of $\mathbf{I}_{k_I}$ and 
$\mathbf{I}^{L} = \prod_1^{z_L} \mathbf{I}_{k_L}$ is the direct product of all $z_L$ identity matrices of rank $k_L$ in the $L$-th shell. 

Considering the translational symmetry of the supercell, after Fourier transforming, 
one can evaluate the averaged Green's function: 
\begin{subequations}
\begin{eqnarray}
\mathbf{G}_{00}(\mathbf{q},\omega) &=& \BRA{F_0} \hbar \left[ \omega \textbf{\~I} - \textbf{\~H}(\mathbf{q}) \right]^{-1} \KET{F_0} \;, \\
\textbf{\~H}(\mathbf{q}) &=&  \mathbf{Q}(\mathbf{0}) - \mathbf{Q}(\mathbf{q})  \;, 
\end{eqnarray}
\end{subequations}
where, in the shell structure form, 
\begin{subequations}
\begin{eqnarray}
\mathbf{Q}(\mathbf{q}) &=& \sum_{I} \mathbf{Q}^I(\mathbf{q}) \;, \\
\mathbf{Q}^I(\mathbf{q}) &=& \mathbf{I}^1 \otimes \cdots \otimes \mathbf{I}^{I-1} 
                             \otimes \mathbf{M}^{I}(\mathbf{q}) \otimes 
                             \mathbf{I}^{I+1} \cdots \!\!, \\
\mathbf{M}^I(\mathbf{q}) &=& \sum_{n=1}^{z_I} \mathbf{M}_n^I(\mathbf{q}) \;, \\
\mathbf{M}_n^I(\mathbf{q}) &=& ( \prod_{1}^{n-1} \otimes \mathbf{I}_{k_I} )
                   \otimes M_{I}^{k_I}(\mathbf{q}) \otimes 
                   ( \prod_{n+1}^{z_I} \otimes \mathbf{I}_{k_I} ) \!, 
\end{eqnarray}
\end{subequations}
where $M_{I}^{k_I}(\mathbf{q}) = M_{I}^{k_I} \gamma_I (\mathbf{q})$ (the proof being given in Appendix). 

Furthermore, the averaged Green function $\EW{\mathbf{G}(\omega)}$ can be evaluated from its momentum representation  
\begin{equation}
\EW{\mathbf{G}(\omega)}=\sum_{\mathbf{q}} \mathbf{G}_{00}(\mathbf{q},\omega) \;,
\end{equation}
where the Lambin-Vigneron algorithm\cite{Lambin84} is used to do summation in $\mathbf{q}$-space over the Brillouin zone.

Then, the magnon spectral function can be expressed as 
\begin{equation}
\label{eq:DOS AB}
  A(\omega) = - \frac{1}{\pi} \text{Im} \EW{\mathbf{G}(\omega)} \;.
\end{equation}
Using the Callen equation, magnetization reads
\begin{equation}
  \label{eq:Sz_all_S}
  \EW{S^z}=\hbar\frac
  {\left(1+S+\Psi\right)\Psi^{2S+1}+
    \left(S-\Psi\right)\left(1+\Psi\right)^{2S+1}}
  {\left(1+\Psi\right)^{2S+1}-\Psi^{2S+1}}
\end{equation}
where the average magnon number $\Psi$ can be calculated by
\begin{equation}
\label{eq:eqfi}
  \Psi=\int_{-\infty}^{+\infty} d\omega
  \frac{A(\omega)}{e^{2\hbar \EW{s^z}\omega/k_B T}-1 } \;.
\end{equation}

Considering $\EW{S^{z}}\rightarrow 0$ in the limit $T\rightarrow T_C$, one can get the transition temperature 
\begin{equation}
\label{eq:eqtc}
k_{B}T_{C}=\frac{2}{3}\hbar S(S+1) 
\left( \int_{-\infty}^{+\infty} d\omega \frac{A(\omega)}{\omega} \right)^{-1} \;.
\end{equation}

\section{Numerical Studies}
\label{sec:Numerical_Studies}

In this section, the influence of dilution and disorder on magnetization and Curie temperature is investigated in diluted spin systems. 
For simplicity, we only consider the case of a three-dimensional system 
on a simple cubic lattice with the lattice constant $a=1$ and an infinitesimal external magnetic field. 
In addition, the shrink of the $\mathbf{q}$-space is included because of the Fourier transforming based on the supercells. 

To study the influence of the range of a ferromagnetic exchange interaction on magnetization, respectively, 
power-law, RKKY and damped-RKKY type exchange interactions are used: 
\begin{equation}
\label{eq:J_R_longrange}
\!J_{ij}(R) = \begin{cases}
J_0 / (\frac{R}{a})^4 \;, \\
J_0 \left[\text{sin}(\frac{R}{a})-(\frac{R}{a})\text{cos}(\frac{R}{a}) \right] / (\frac{R}{a})^{4} \;, \\
J_0 e^{-\frac{R}{R_0}} \left[\text{sin}(\frac{R}{a})-(\frac{R}{a})\text{cos}(\frac{R}{a}) \right] / (\frac{R}{a})^4 \!,
\end{cases}
\end{equation}
where $a$ is the lattice constant, $J_0$ is the effective nearest-neighbor interaction strength and 
$R$ is the distance between two magnetic ions. 
If the damping factor $R_0 \rightarrow \infty$, the damped-RKKY exchange interaction is equal to the RKKY exchange interaction. 
In order to calculate practicably, one needs to do cutting about the exchange interaction at a distance $R_{cut-off}$. 
Considering the Fourier transformation based on the supercells 
and comparing the calculations of the \textit{diluted and ordered} case and of the \textit{diluted and disordered} case, 
the \textit{diluted and ordered} case is used to decide the shell structure of systems. 
If the distance between one supercell and the central supercell is larger than $R_{cut-off}$, 
the exchange interactions will be considered as zero. 
It should be mentioned that the equation (\ref{eq:J_R_longrange}) is used just for illustration of the present approach, 
although the neglected effect of disorder on the value of exchange integrals\cite{Bergqvist05,Sato04,Silva05} should be included. 
In addition, superexchange effects and the influence of virtual bound states on the value of exchange integrals
\cite{Bouzerar06,QZhang86} are also neglected. 
\begin{figure}[tb]
\vspace*{1.0cm}
\centerline{\includegraphics[width=0.8\linewidth]{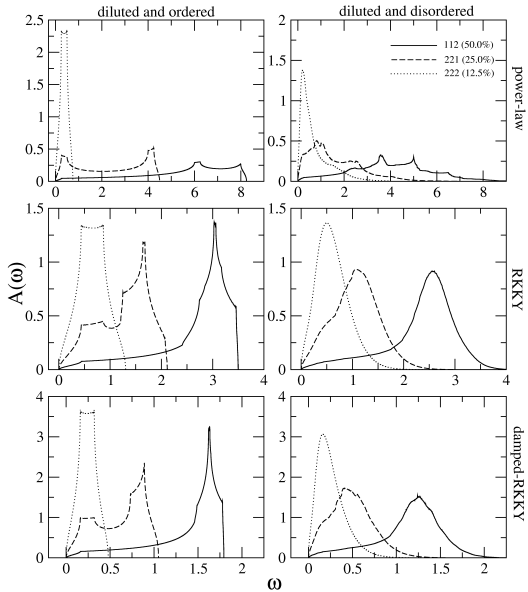}}
\caption{
\label{fig:dos}
The magnon spectral density for $R_{cut-off}/a = 2.0$ on an sc lattice for various 
concentrations $x$ (that correspond to the different sizes of the supercells) 
for the \textit{diluted and ordered} case and the \textit{diluted and disordered} case with 
the effective nearest-neighbor integral $J_0 = 1.0$, $S=2.5$ and the damping factor $R_0 = 2.0 a$. 
}
\end{figure}

Fig. \ref{fig:dos} shows the magnon spectral density of the \textit{diluted and ordered} case 
and the \textit{diluted and disordered} case 
for several concentrations $x$ that correspond to the different sizes of the supercells. 
Because of the cut-off distance $R_{cut-off} = 2.0 a$, the exchange parameters 
between the nearest-neighbor supercells are only included for the $\{2 \times 2 \times 2\}$ supercells' structure  
that corresponds to the concentration $x=12.5\%$. 
That is the reason why the calculation of the $\{2 \times 2 \times 2\}$ structure in the \textit{diluted and ordered} case 
shows the well-known symmetric shape of a simple cubic density of states except for the change of the bandwidth. 
For the $\{1 \times 1 \times 2\}$ and $\{2 \times 2 \times 1\}$ supercells' structures, in the \textit{diluted and ordered} case, 
there are more than one peak because of including the exchange interactions 
belonging to next nearest-neighbor(nnn) or other longer-range supercells. 
In addition, there is one peak in the $\{1 \times 1 \times 2\}$ and $\{2 \times 2 \times 1\}$ structures 
being in the same position as that in the $\{2 \times 2 \times 2\}$ structure. 
It means all the magnetic ions and the exchange interactions in the $\{2 \times 2 \times 2\}$ structure 
are also included in the $\{1 \times 1 \times 2\}$ and $\{2 \times 2 \times 1\}$ structures. 
It should be noted here that the long-range exchange interactions between magnetic ions are included 
in the nearest-neighbor supercells because the exchange interactions are based on the positions of magnetic ions. 
Fig. \ref{fig:dos} also shows that the peaks of the magnon spectral density 
move toward lower energies for decreasing concentrations of the magnetic ions. 
It shows that dilution increases the magnon spectral density for lower energies 
at the cost of the magnon spectral density at higher energies. 

Disorder also enhances the low energy part of the magnon spectral density, 
especially for the $\{1 \times 1 \times 2\}$ and $\{2 \times 2 \times 1\}$ structures, 
if one compares the \textit{diluted and ordered} case and the \textit{diluted and disordered} case in Fig. \ref{fig:dos}. 
However, it is clearly shown in the figures that the influence of dilution is more important than that of disorder. 
Furthermore, the obviously separated peaks in the \textit{diluted and ordered} case 
disappear in the \textit{diluted and disordered} case 
because the random position distribution of the magnetic ions leads to a dispersion of the exchange integrals, 
while the exchange integral between two supercells is a single value in the \textit{diluted and ordered} case. 
At the same time, it is clear that the dispersion of the exchange integrals 
leads to the long tails that extend to high energy magnons' area, 
especially for the $\{2 \times 2 \times 2\}$ structure in the panel of power-law interactions. 

It is quite interesting to note that the damping factor also increases the magnon spectral density for lower energies. 
From the expression of the damped-RKKY exchange integrals, 
one finds that the exponential damping will decrease the strength of the short-range interaction less than that of the long-range one. 
Compared with the RKKY exchange interaction, it leads to the clearer separation of the exchange integral distribution 
for the damped-RKKY exchange interaction in the \textit{diluted and ordered} case. 
It is especially clear for the $\{2 \times 2 \times 1\}$ structure. 
But in the \textit{diluted and disordered} case, 
the changing of the separation is not so obvious due to the dispersions of the exchange integrals. 

\begin{figure}[tb]
\vspace*{1.0cm}
\centerline{\includegraphics[width=0.8\linewidth]{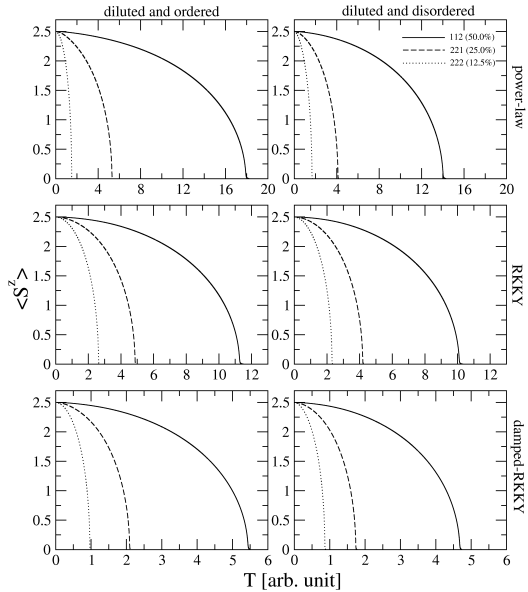}}
\caption{
\label{fig:s_z}
Magnetization $\EW{S^z}$ as function of temperature $T$ for $R_{cut-off}/a = 2.0$ on an sc lattice for various 
concentrations for the \textit{diluted and ordered} case and the \textit{diluted and disordered} case with 
$J_0 = 1.0$, $S=2.5$ and $R_0 = 2.0 a$. 
}
\end{figure}
The temperature dependence of magnetization, which is calculated from 
the magnon spectral density in Fig. \ref{fig:dos}, is shown in Fig. \ref{fig:s_z}. 
The figures show that all curves of magnetization decreases to zero monotonically and smoothly 
with increasing temperature from $T=0$ to $T=T_c$. 
From the equation (\ref{eq:eqtc}), one knows that increasing magnon spectral density at lower energies 
will lead to decrease the Curie temperature. 
It is clearly shown by the influence of dilution on Curie temperature 
because dilution strongly shifts the magnon spectral density to lower energies. 
Dilution increases the distance between two magnetic ions and further 
decreases the strength of the exchange integrals. 
In other words, the influence of dilution on Curie temperature mainly comes from decreasing of the effective exchange interactions. 

Disorder also influences Curie temperature $T_c$, but not so important as dilution. 
The random distribution of the magnetic ions leads to the dispersions of the exchange integrals. 
That increases the lower energy part of magnon spectral density and is the reason why disorder lowers the Curie temperature. 
It is especially clear for the $\{1 \times 1 \times 2\}$ and $\{2 \times 2 \times 1\}$ structures. 
However, for the $\{2 \times 2 \times 2\}$ structure of power-law interactions, 
Curie temperature $T_c$ is slightly higher in the disordered case than that in the ordered case 
because the dispersions also increases the high energy part of magnon spectral density. 
The final effect of disorder on $T_c$ is the results of competition between the high energy magnons and the low energy magnons. 
If one uses an average value or a special value to substitute the dispersion of the exchange integrals, 
just like the mean-field approximation ($k_B T_c \sim \frac{2}{3N}\sum_{\mathbf{q}} J^{eff}(\mathbf{q})$) 
or the \textit{diluted and ordered} case in our calculation, 
the influence of the dispersions of the exchange integrals will be washed out. 

The results show that the Curie temperature is mainly influenced by dilution or the effective exchange interactions. 
It may explain the following fact. 
For a long time, it was rather surprising that the models that neglect the disorder and treat the exchange interaction in 
MFA provide a Curie temperature in a good agreement with the experiment. 
Although the positions of magnetic ions is randomly distributed, 
one may get a reasonably calculated Curie temperature, which is near to the experimental Curie temperature, 
by adjusting suitably the effective exchange integrals. 
Moreover, given the level of dilution of the magnetic ions, the results suggest that 
the way to get high Curie temperature DMS materials is increasing the effective exchange integrals $J_0$. 

Comparing the calculation of RKKY and damped-RKKY, 
we find there is a big influence of the exponential damping on Curie temperature, 
which is even a more important factor than the position disorder of magnetic ions for $R_0=2.0 a$ in our calculation. 
The reasons to introduce the exponential damping is due to substitutional disorder or 
to the half-metallic character of (Ga,Mn)As and (Ge,Mn) alloy, etc\cite{Kudrnovsky04,Fiete05,Brey03}. 
In our theory, only the latter is meant because the positional disorder is 
already explicitly taken into account in our calculation. 
The obvious influence of damping on Curie temperature suggests that the damping factor should be chosen carefully 
if one only wants to incorporate the damping exponential item to include the positional disorder influence, 
especially in model calculation. 

\section{Summary}
\label{sec:Summary}

The aim of this article is to study the influence of dilution and disorder 
on the ferromagnetic properties of diluted spin systems. 
By combining the supercell approach and the ASF, 
we study theoretically the influence of disorder on magnetization and Curie temperature 
in diluted spin systems on a disordered Heisenberg model. 
Firstly, the size of the supercell is used to determine the concentration of magnetic ions 
because there is only one magnetic ion per supercell in our calculations. 
In order to investigate the influence of dilution and disorder on Curie temperature, 
the positions of magnetic ions in the supercells can be arranged in two ways: 
(i) the \textit{diluted and ordered} case in which the magnetic ion is only able to occupy the same lattice within each supercell and 
(ii) the \textit{diluted and disordered} case in which the magnetic ion is able to occupy any lattice point in the supercell. 
For the \textit{diluted and ordered} case and the concentrated case, 
the equations of motion of the magnon Green's function, which is decoupled by making Tyablikov approximation, 
can be solved directly from Fourier transformation because of the translational symmetry of the supercells. 
For the \textit{diluted and disordered} case, the effective Heisenberg exchange integrals, 
which are assumed to be a function of distance only, are random variables 
because the positions of two magnetic ions in the supercells is randomly distributed. 
By using ASF, the random exchange integrals between two supercells are extended to matrices, 
therewith restalling translational symmetry, in order to calculate the averaged magnon spectral density. 
Then, the obtained averaged spectral densities have been used to 
calculate the temperature dependence of magnetization and the Curie temperature. 
Significantly, the long-range exchange integrals are included and 
all the calculations correspond to an infinite system and there is no cutting in the real space. 

The resulting theory is then solved numerically for a simple cubic system. 
The long-range ferromagnetic exchange integrals including power-law decaying, RKKY type and damped-RKKY type 
are used to calculate the temperature dependence of magnetization and Curie temperature. 
The results show that dilution and the damping exponential item increase the magnon spectral density 
for lower energies at the cost of the magnon spectral density at higher energies. 
That leads to a decrease of the Curie temperature of spin systems. 
The influence of dilution on Curie temperature mainly comes from decreasing the effective exchange interactions. 
The effect of disorder on Curie temperature is more complicated. 
The random distribution of the magnetic ions leads to dispersions of the exchange integrals, 
which mainly decrease Curie temperature but sometimes increase $T_c$ in the concentration area of our calculations. 
The role of the exponential damping of distance is obvious and should be set carefully in any phenomenological model. 
From our calculations, to attack the problem of high temperature ferromagnetic DMS materials, 
the effective way is increasing the effective exchange integrals.


\begin{acknowledgments}
We thank Prof. Avinash Singh for helpful discussions. 
G.X.T is supported by the State Scholarship Programs of China Scholarship Council, 
the National Natural Science Foundation of China under Grant No. 50375040 and 
Foundation of HIT Grant No. HIT.MD2002.16.
\end{acknowledgments}

\section{Appendix}
\label{sec:Appendix}

For $\mathbf{R}_{ij}$, if the index $i$ refers to the $I$-th shell and $j$ refers to the $j$-th supercells in $I$-th shell, 
we write Fourier transformation of $\mathbf{J}_{ij}=\mathbf{J}(R_{ij})$ as: 
\begin{eqnarray*}
\mathbf{Q}(\mathbf{q}) &=& \sum_{R_{ij}} e^{-i\mathbf{R}_{ij} \cdot \mathbf{q}} \mathbf{J}(R_{ij}) \\
                       &=& \left[ e^{-i\mathbf{R}_{1 1}   \cdot \mathbf{q}} \mathbf{J}(R_{1 1}) + \cdots + 
		                  e^{-i\mathbf{R}_{1 z_1} \cdot \mathbf{q}} \mathbf{J}(R_{1 z_1}) \right] \\
                       &+& \left[ e^{-i\mathbf{R}_{2 1}   \cdot \mathbf{q}} \mathbf{J}(R_{2 1}) + \cdots + 
		                  e^{-i\mathbf{R}_{2 z_2} \cdot \mathbf{q}} \mathbf{J}(R_{2 z_2}) \right] \\
		       &+& \cdots \\
                       &+& \left[ e^{-i\mathbf{R}_{I 1}   \cdot \mathbf{q}} \mathbf{J}(R_{I 1}) + \cdots +
		                  e^{-i\mathbf{R}_{I z_I} \cdot \mathbf{q}} \mathbf{J}(R_{I z_I}) \right] \\
		       &+& \cdots \\
		       &=& \sum_I \mathbf{Q}^I(\mathbf{q}) 
\end{eqnarray*}
where $z_I$ is the total number of the supercells in the $I$-th shell and 
\begin{eqnarray*}
\mathbf{Q}^I(\mathbf{q}) &=& \sum_{n=1}^{z_I} e^{-i\mathbf{R}_{In} \cdot \mathbf{q}} \mathbf{J}(R_{In}) \\
                         &=& I^1 \otimes \cdots \otimes I^{I-1} \otimes \mathbf{M}^I(\mathbf{q}) \otimes I^{I+1} \otimes \cdots \;, \\
\mathbf{M}^I(\mathbf{q}) &=& \sum_{n=1}^{z_I} e^{-i\mathbf{R}_{In} \cdot \mathbf{q}} \mathbf{M}^I \;, 
\end{eqnarray*}
If we refer the $n$-th supercells to that of the central position $\mathbf{R}_{In}$, i.e., 
\begin{eqnarray*}
\{ 1,2, \cdots, (z_I-1), z_I \} \rightarrow 
         \{ \mathbf{R}_{I1}, \mathbf{R}_{I2}, \cdots, \mathbf{R}_{I(z_I-1)}, \mathbf{R}_{Iz_I} \} \;, 
\end{eqnarray*}
$\mathbf{M}^I(\mathbf{q})$ can be expressed as 
\begin{eqnarray*}
\mathbf{M}^I(\mathbf{q})
 &=&e^{-i\mathbf{q}\cdot \mathbf{R}_{I 1}  }(M_{I}^{k_I} \otimes I_{k_I} \otimes I_{k_I} \otimes \cdots \otimes I_{k_I})\\
 &+&e^{-i\mathbf{q}\cdot \mathbf{R}_{I 2}  }(I_{k_I} \otimes M_{I}^{k_I} \otimes I_{k_I} \otimes \cdots \otimes I_{k_I})\\
 &+& \cdots \\
 &+&e^{-i\mathbf{q}\cdot \mathbf{R}_{I z_I}}(I_{k_I} \otimes I_{k_I} \otimes \cdots \otimes I_{k_I} \otimes M_{I}^{k_I})
\end{eqnarray*}
In addition, for 
\begin{eqnarray*}
\{ 1,2, \cdots, z_I \} &\rightarrow&
         \{ \mathbf{R}_{I2}, \mathbf{R}_{I3}, \cdots, \mathbf{R}_{I z_I}, \mathbf{R}_{I1} \} \;, \\
	               &\vdots& \\
\{ 1,2, \cdots, z_I \} &\rightarrow&
         \{ \mathbf{R}_{Iz_I}, \mathbf{R}_{I1}, \cdots, \mathbf{R}_{I(z_I-2)}, \mathbf{R}_{I(z_I-1)} \} \;, 
\end{eqnarray*}
correspondingly, one can write 
\begin{eqnarray*}
\mathbf{M}^I(\mathbf{q})
 &=&e^{-i\mathbf{q}\cdot \mathbf{R}_{I 2}   }(M_{I}^{k_I} \otimes I_{k_I} \otimes I_{k_I} \otimes \cdots \otimes I_{k_I})\\
 &+&e^{-i\mathbf{q}\cdot \mathbf{R}_{I 3}   }(I_{k_I} \otimes M_{I}^{k_I} \otimes I_{k_I} \otimes \cdots \otimes I_{k_I})\\
 &+&\cdots \\
 &+&e^{-i\mathbf{q}\cdot \mathbf{R}_{I 1} }(I_{k_I} \otimes I_{k_I} \otimes \cdots \otimes I_{k_I} \otimes M_{I}^{k_I}) \;, 
\end{eqnarray*}
\begin{eqnarray*}
 &\vdots& \\
\mathbf{M}^I(\mathbf{q})
 &=&e^{-i\mathbf{q}\cdot \mathbf{R}_{I z_I}   }(M_{I}^{k_I}\otimes I_{k_I}\otimes I_{k_I} \otimes \cdots \otimes I_{k_I})\\
 &+&e^{-i\mathbf{q}\cdot \mathbf{R}_{I 1}     }(I_{k_I}\otimes M_{I}^{k_I}\otimes I_{k_I} \otimes \cdots \otimes I_{k_I})\\
 &+&\cdots \\
 &+&e^{-i\mathbf{q}\cdot \mathbf{R}_{I(z_I-1)}}(I_{k_I}\otimes I_{k_I}\otimes \cdots \otimes I_{k_I} \otimes M_{I}^{k_I})\;. 
\end{eqnarray*}
To sum $z_I$ expressions of $M^I(\mathbf{q})$ and divided by $z_I$, one can get 
\begin{eqnarray*}
\mathbf{M}^I(\mathbf{q})   &=& \sum_{n=1}^{z_I} \mathbf{M}_n^I(\mathbf{q}) \;, \\
\mathbf{M}_n^I(\mathbf{q}) &=& ( \prod_{1}^{n-1} \otimes \mathbf{I}_{k_I} ) 
                                 \otimes M_{I}^{k_I}(\mathbf{q}) \otimes 
		               ( \prod_{n+1}^{z_I} \otimes \mathbf{I}_{k_I} ) \;, 
\end{eqnarray*}
where $M_{I}^{k_I}(\mathbf{q}) = M_{I}^{k_I} \gamma_I (\mathbf{q})$ and 
$\gamma_I(\mathbf{q}) = \frac{1}{z_I} \sum_{n=1}^{z_I} e^{-i \mathbf{R}_{In} \cdot \mathbf{q} }$. 




\end{document}